# The status of the search for low mass WIMPs: 2013

David B. Cline
*Astroparticle Physics, UCLA*

## Abstract

Using information from a recent dark matter symposium at Marina del Rey and from various publications in 2012 and 2013, we discuss the most recent evidence and constraints on low mass WIMPs. There are now five separate experimental limits on such WIMPs, including a new paper on the XENON100 225 day exposure. There are very different experimental methods with different backgrounds that comprise this limit. We speculate on the possible sources of the reported low mass WIMP signals and background. We present recent arguments concerning DAMA that show the possible DM claims are likely misleading. We discuss the new CDMS claims for a signal and question the very low ionization in these events. We also discuss an analysis of XENON 100 data that uses information theory that further excludes the CDMS results.

**Introduction**

With the discovery of a 125 GeV particle by CMS [1] and Atlas [2] that is widely believed to be the Higgs boson, various models of supersymmetric WIMPs increase the expected mass to the 500 GeV or greater and cross-sections to between $10^{-45}$ to $10^{-47}$ $cm^2$ . Only the lower edge of this region has been explored by XENON 100s: 225 day exposure [3]. Our previous work described the search for low mass WIMPs [4].

The likelihood of a supersymmetric low mass WIMP from the theory is very remote. Nevertheless claims from DAMA, CoGeNT, and CRESST have not been withdrawn. This is an unfortunate problem in the worldwide search for dark matter particles. At the recent Dark Matter symposium at Marina del Rey there was very strong evidence put forth to limit the possibility of low mass WIMPs [5]. In particular the null CDMS II search for annual variation in the low mass range coupled with the latest XENON100, 225 day exposure strongly constrains the low mass WIMP hypothesis [6]. Recently CDMS has claimed a low-mass signal.

In this paper we will present all the current evidence for the low mass WIMP search.

## 1. Summary of world limit on low mass WIMP signals

In Figure 1 we show a summary of the current limits on the low mass WIMP region [3]. We note that the CDMS II, Simple, XENON10 limits come from very different methods:

- CDMS II Ultra cold Ge detector

- XENON10 Use of the $S_2$ only signal from a special low threshold run of XENON 10
- SIMPLE Use of heated droplets
- XENON100 Use of Xenon detector using (2 experiments) traditional methods of $S_1$ and $S_2$

Because the claimed cross-section is so large, these methods are all very robust.

The limits from XENON100 deserve a special discussion [3]. Both the 100 day XENON100 exposure and the more recent 225 day exposure are inconsistent with a low mass WIMP to the 90 percent confidence level. These data are totally independent and not summed in Figure 1. One could assume that the new 225-day data logically reinforce the 100-day limit. There are then five limits: Simple, XENON10 (S2), CDMS II, XENON100, 100 days, and XENON 225-day limits. All are independent and are 90 percent confidence level null limits. We note that the DAMA results are reported as 3σ limits (see Reference 4 for references).

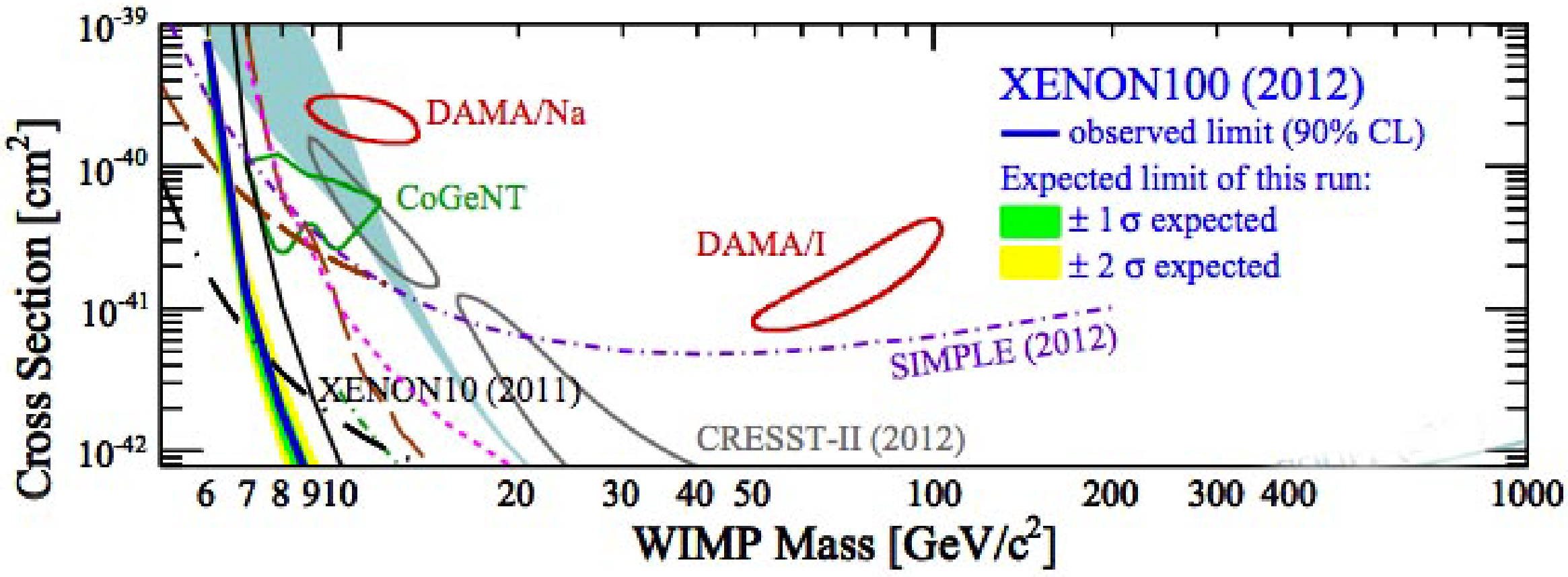

Figure 1. An enlargement of the low mass scale of WIMP searches from the recent XENON100 225 day paper (E. Aprile et al, "Dark Matter Results from 225 Live Days of XENON100 Data," http://arxiv.org/abs/1207.5988).

## 2. The use of the $S_2$ signal from a Xenon on Argon dark matter detector

Normally the $S_2$ signal used to carry out discrimination of a WIMP signal is from an EM background. However, as shown by the ZEPLIN II group who first measured $S_2$ experimentally [7] and from experiments by the UCLA Torino team in the 1990s, this parameter is very robust. When a particle hits a large atom like Xenon the outer electrons are easily stripped off. With an electric field applied the free electrons drift to a gas system that provides amplification. Typically one electron from the vertex can yield 20 to 30 photoelectrons in the experiment's PMTs, giving $S_2$. $S_2$ is usually used to trigger the detector. It was recognized early on that the $S_2$ signal could be used to measure energy in low energy events [8]. Recently a UCLA study has shown a way to analyze data using the $S_2$ signal [9]. See also the work of P. Sorensen and the XENON10 group [10]. The essence of this subject is that very low energy recoils can be detected with the $S_2$ signal, making this appropriate to search for very low mass WIMP signals. The $S_2$ signal has a lower threshold than any other current dark matter detector.

There are two choices:

(a) Use a small $S_1$ signal to determine the position of the event in the detector and use $S_2$ to measure the energy [9];
(b) Use the diffusion of the $S_2$ signal to measure Z (upward) position and determine x, y (z) for the event [10].

The (b) method has been used to determine the limit on low mass WIMPs shown in Figure 2 showing a very robust limit (also shown in Figure 1) [10].

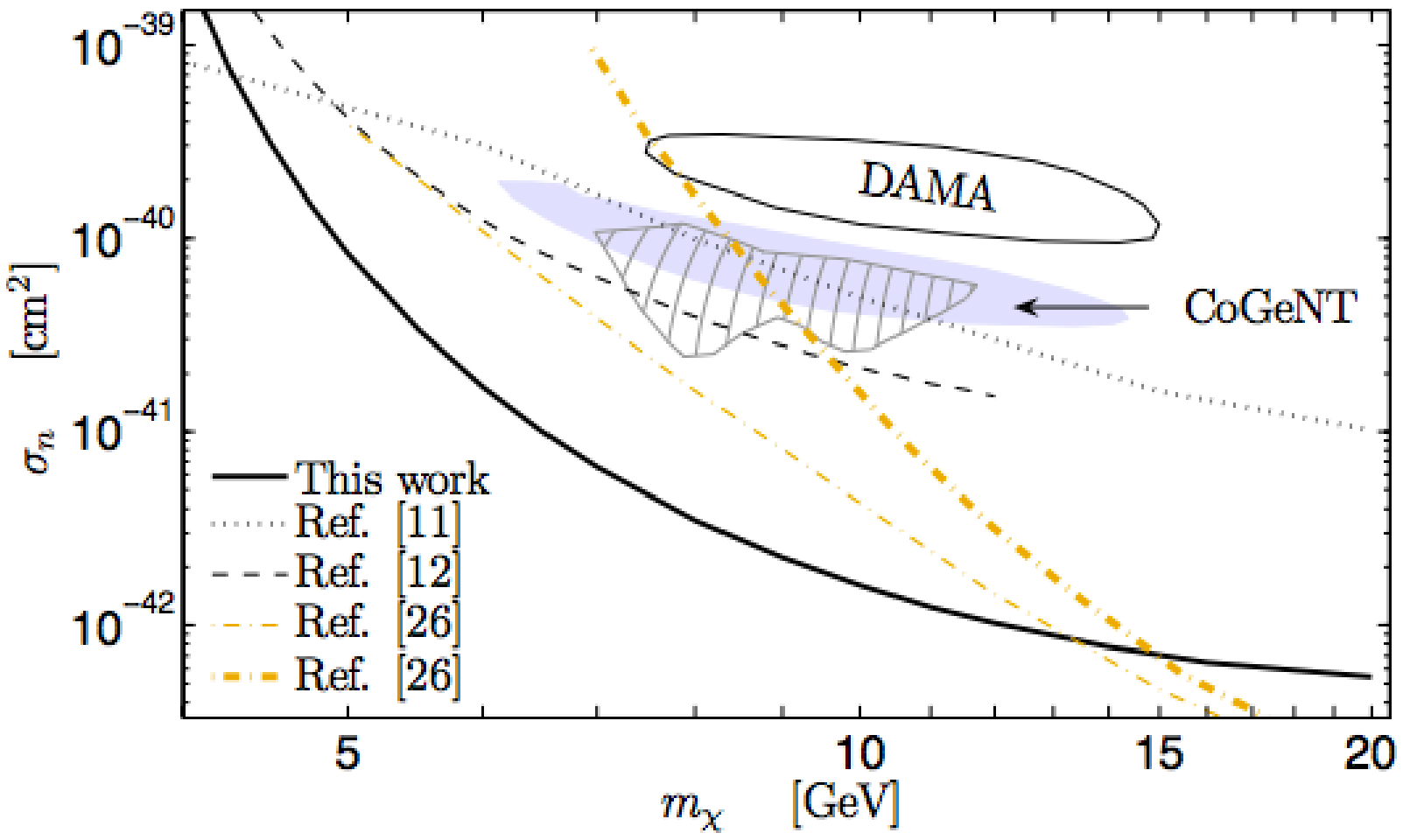

Figure 2. Curves indicate 90% C.L. exclusion limits on spin-independent $\sigma_n$ for elastic dark matter scattering, obtained by CDMS (dotted [11], and dashed [12]) and XENON100 (dash-dot [26]). The region consistent with assumption of a positive detection by CoGeNT is shown (hatched) [2], and (shaded)[4]; the latter assumes a 30% exponential background. Also shown is the 3σ allowed region for the DAMA annual modulation signal (solid contour) [40] (see Reference 10).

### 3. The Search for Annual Signal Variation with the CDMS II

This CDMS II result is remarkable since not even the expected Radon annual variation is observed. Once a real WIMP signal is observed the observation of annual signal variation is a powerful method to prove that WIMPs have been discovered. At the recent Marina Del Rey Dark Matter conference the CDMS II group presented a search for the Annual Signal Variation observed by the Cogent experiment (recall that CDMS II is a 5kg detector and Cogent is 300 grams). In Figure 4 we show the CDMS II results.

### 4. Recent studies of the effect of K(40) in the DAMA experiment

In reference 11 it is shown that the bulk of the singles signal in DAMA is due to radioactive background. Now a new study (see Fig. 3) shows that less than 0.14 Cpd (Fig. 3) can at most be due to WIMPs. This means that the annual variation of the possible WIMPs signal would have to exceed 20%, which is outside any DM model. Reference 12 gives the results presented in Fig. 3. The excellent agreement with Ref. 11 and the excellent fit strongly suggest this is little or no WIMP signal in the data.

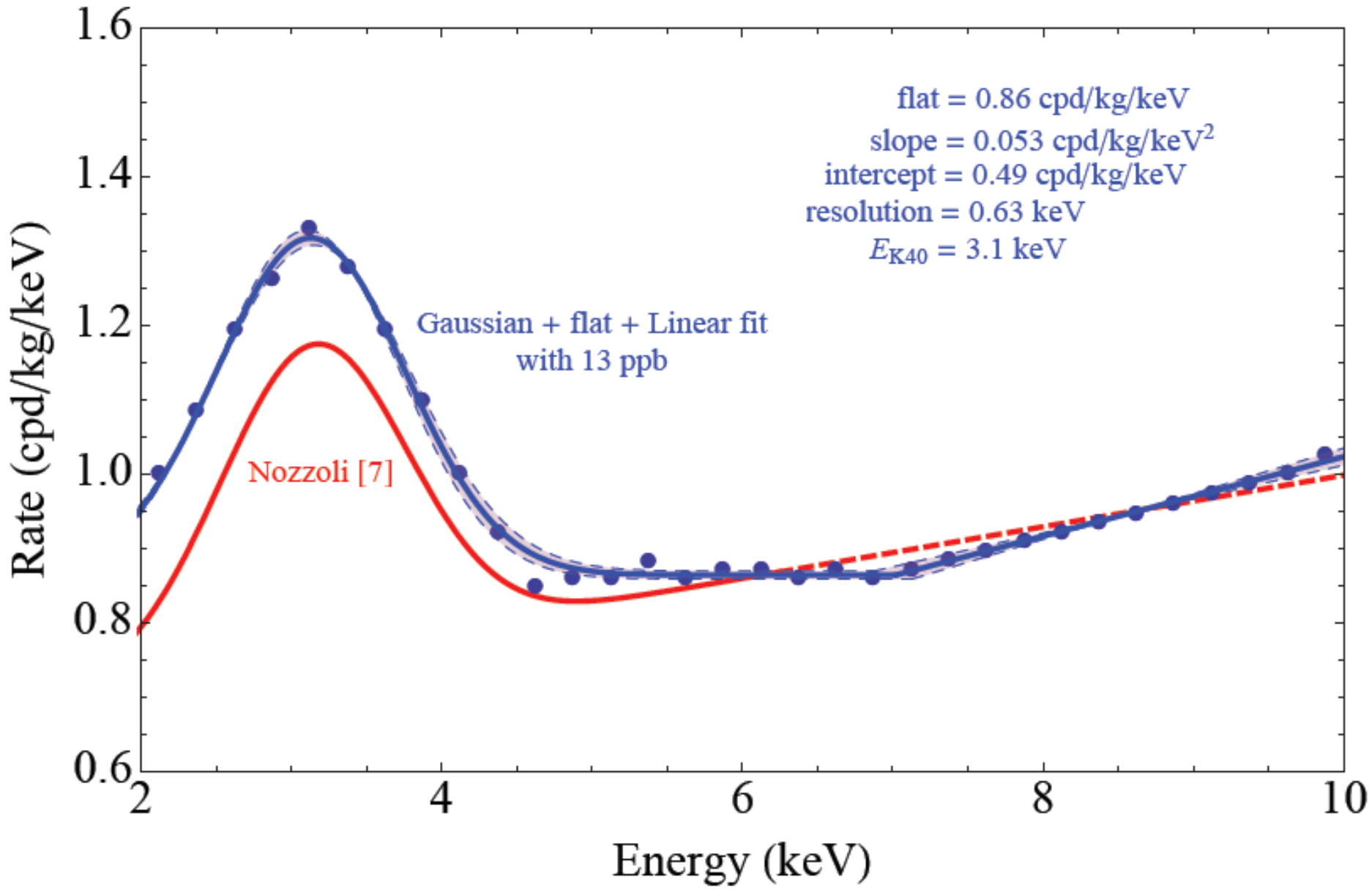

Figure 3. This figure shows the Dama/Libra (dots) and a fit to the data with the correct K (40) and the background from Ref. 11. There is also an estimate by DAMA (red). Even under these circumstances the amount of possible WIMP signal is very low. (Josef Pradler et al, “A reply to criticism of our work (arXiv:1210.5501) by the DAMA collaboration,” http://arxiv.org/abs/1210.7548) [12]

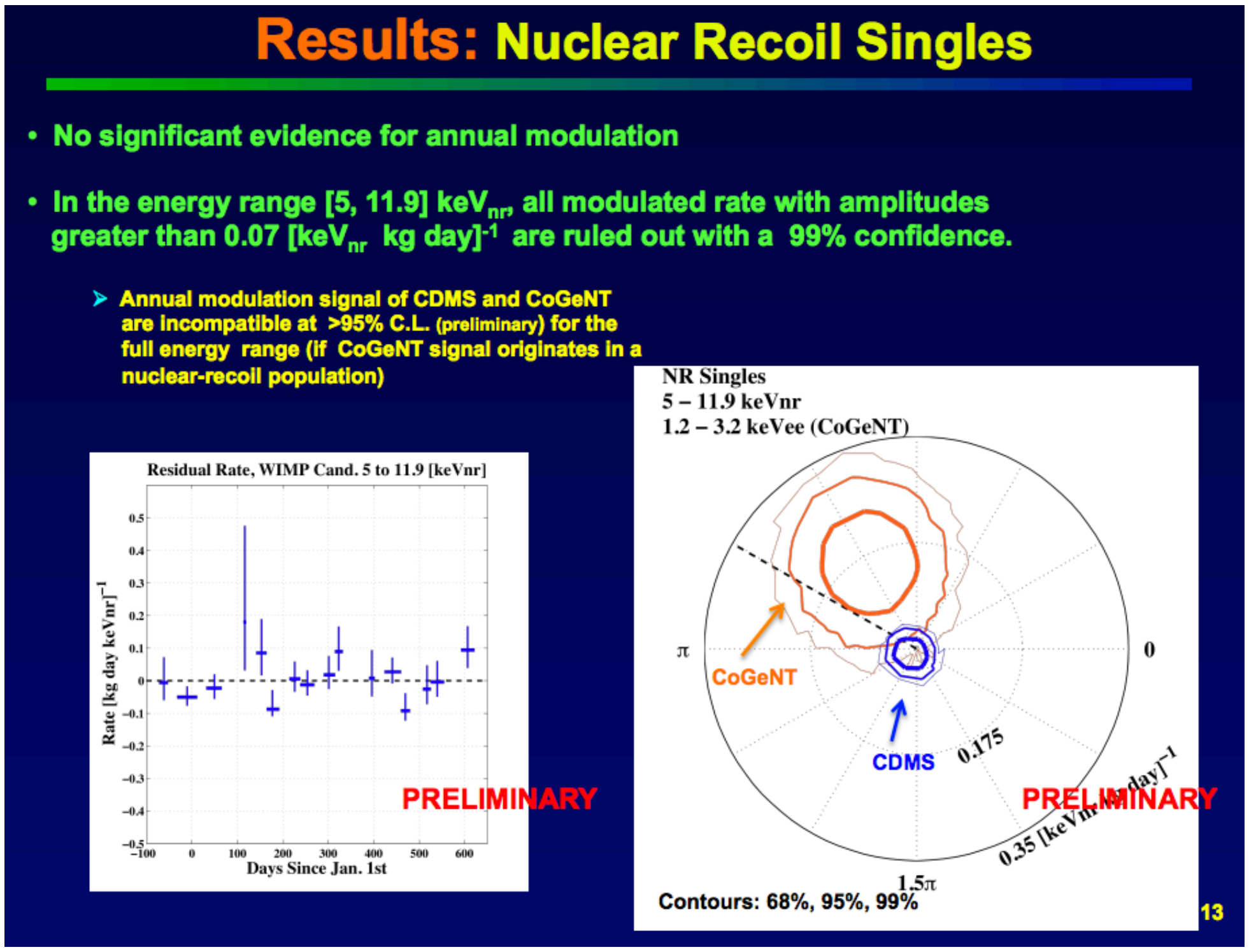

Figure 4. The CDMS group has searched for a low energy signal using the low noise components of the detector as shown in Figure 5. These limits are also shown in Figure 5 (from the CDMS II talk in Reference 5).

In Figure 5 we show the current limits from CDMS II on the low mass region. This analysis used low noise sensors on CDMS II to set this impressive limit [6]. These limits are also shown in Figure 1 and exclude even an enlarged region for DAMA and CoGeNT signals.

## 5. Neutron signals underground

It is well known that the neutron flux in underground labs has an annual variation. This is likely due to the amount of water or snow in the over burden. In the winter the water absorbs neutrons, in the summer much less so. The ICARUS group measured the LGNS neutron flux as shown in Figure 6. Note that this annual variation fits the DAMA data. DAMA is also at the LGNS. J. Ralston took the ICARUS results and extrapolated over the entire DAMA region (Figure 6) (this is not a fit). Note the excellent agreement with the data. We are not claiming that neutrons make the signal in DAMA, only that there are underground sources that seem to fit the same annual variations than one not due to WIMPs.

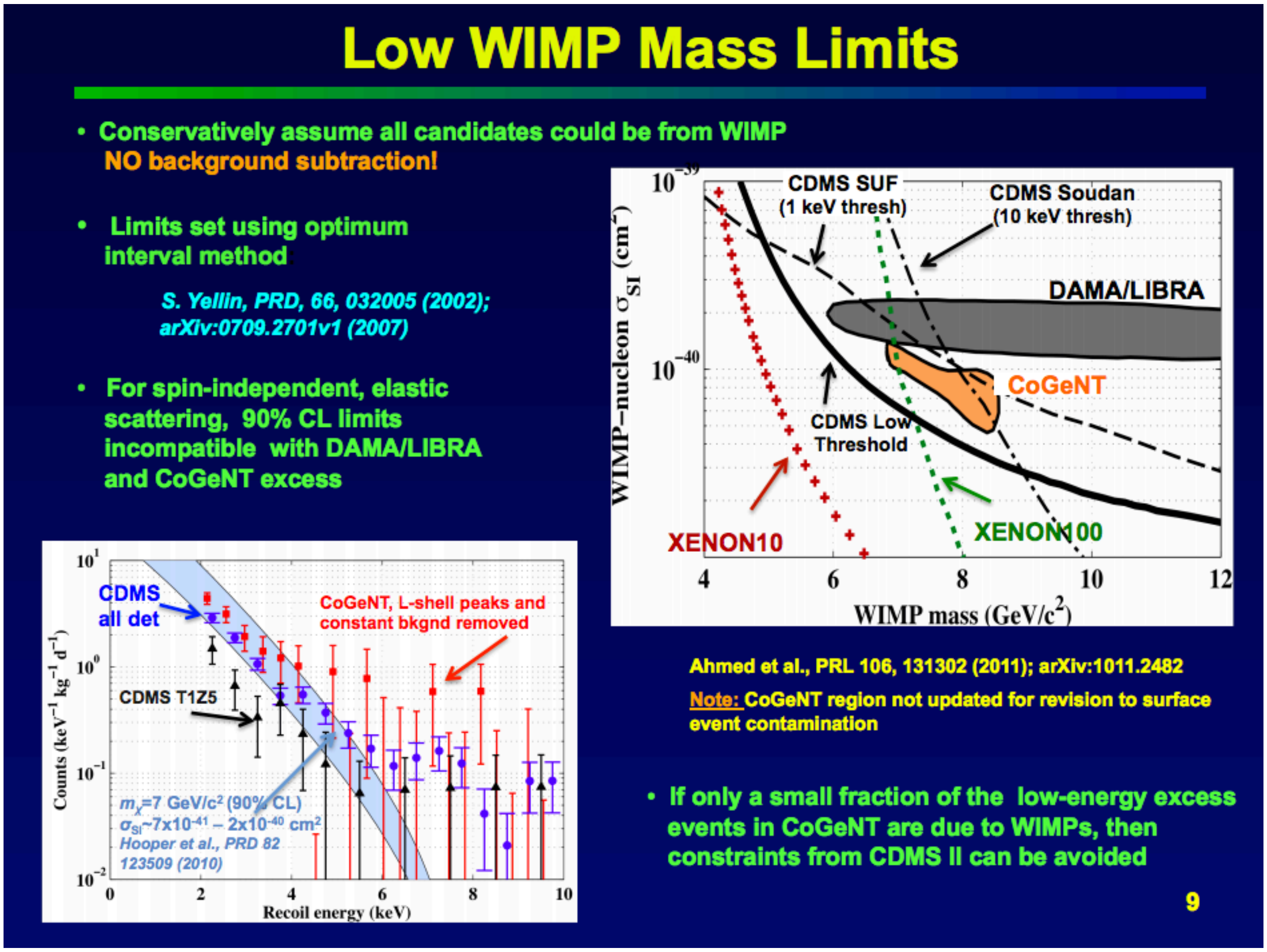

Figure 5. Limits on the low mass WIMP signal from a presentation by the CDMS II group at the Marina del Rey symposium February 2012.

## 6. Signals of Annual Variation underground and DAMA

There are several processes that cause annual variation of processes underground that are similar to the DAMA results.

1. *Radon abundance*
   -Has a clear annual increase in the summer and decrease in the winter seen in all underground laboratories
2. *Variation of neutron flux*
   -In Figure 5 we show neutron intensity data from ICARUS expanded and compared with the DAMA results. All underground laboratories see a neutron flux annual variation.
3. *The annual variation of cosmic muons as compared with DAMA data (Figure 4)*
   -In Figure 7 we show the LVD muon data and compare with the DAMA results (as discussed in Section 4). We do not claim a good fit but there is a general agreement.

For all we know DAMA may be seeing a combination of such effects and the phase they observe would be a mixture of these events. Until we identify the actual source of the signals we will not know the actual phase to predict.

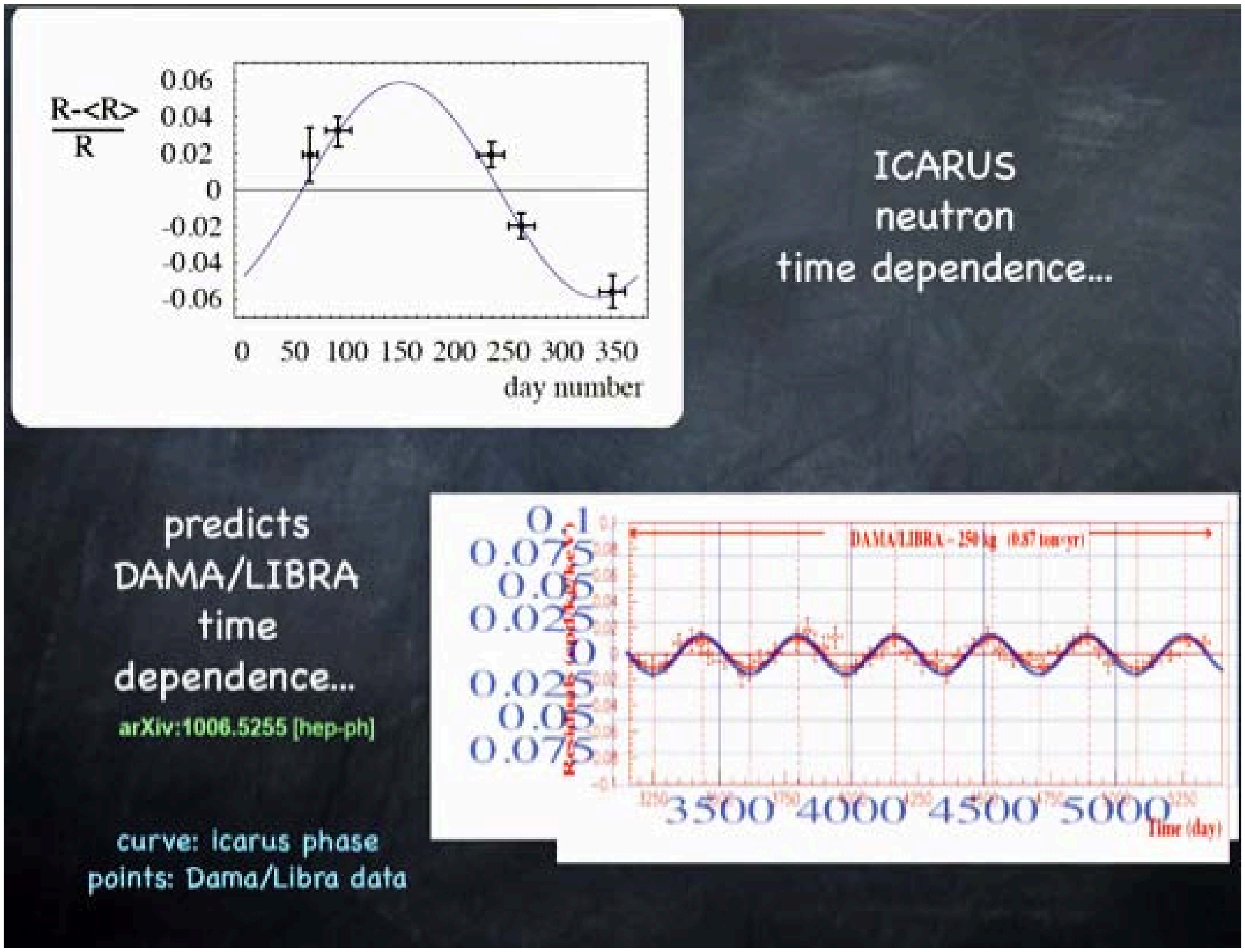

Figure 6. A study of neutron events at the LNGS by the ICARUS group extrapolated to the DAMA results by Ralston (arXiv 1006.5255).

## 7. Overview

The search for low mass dark matter has been mostly negative. While several signals (DAMA, CoGeNT, CRESS II) suggest low mass WIMPs, there are very strong experimental constraints on these signals.

1. 5 separate experiments do not see a nuclear recoil signal consistent with a low mass WIMP (Figure 1).
2. The specific search by CDMS II for annual variation or a direct signal for either CoGeNT or DAMA is null even if the DAMA region is greatly expanded (Figures 4 and 5).
3. While not directly discussed in this paper, the direct comparison of the singles rates in DAMA with a carefully determined radioactive background finds no evidence for a WIMP excess on the data [11]. A similar study has been carried out by Peter Smith at UCLA (unpublished). The result shown in Fig. 3 indicates that very little WIMP production is observed by DAMA.
4. There are several experimental sources of annual variation background that can cause signals in underground detectors. The neutron background measured by ICARUS seems to give a similar signal but others such as Radon and muon also give annual variation. These backgrounds are observed at all underground laboratories and have a simple explanation such as the water load charges in the overburden or the change in density of the upper atmosphere.

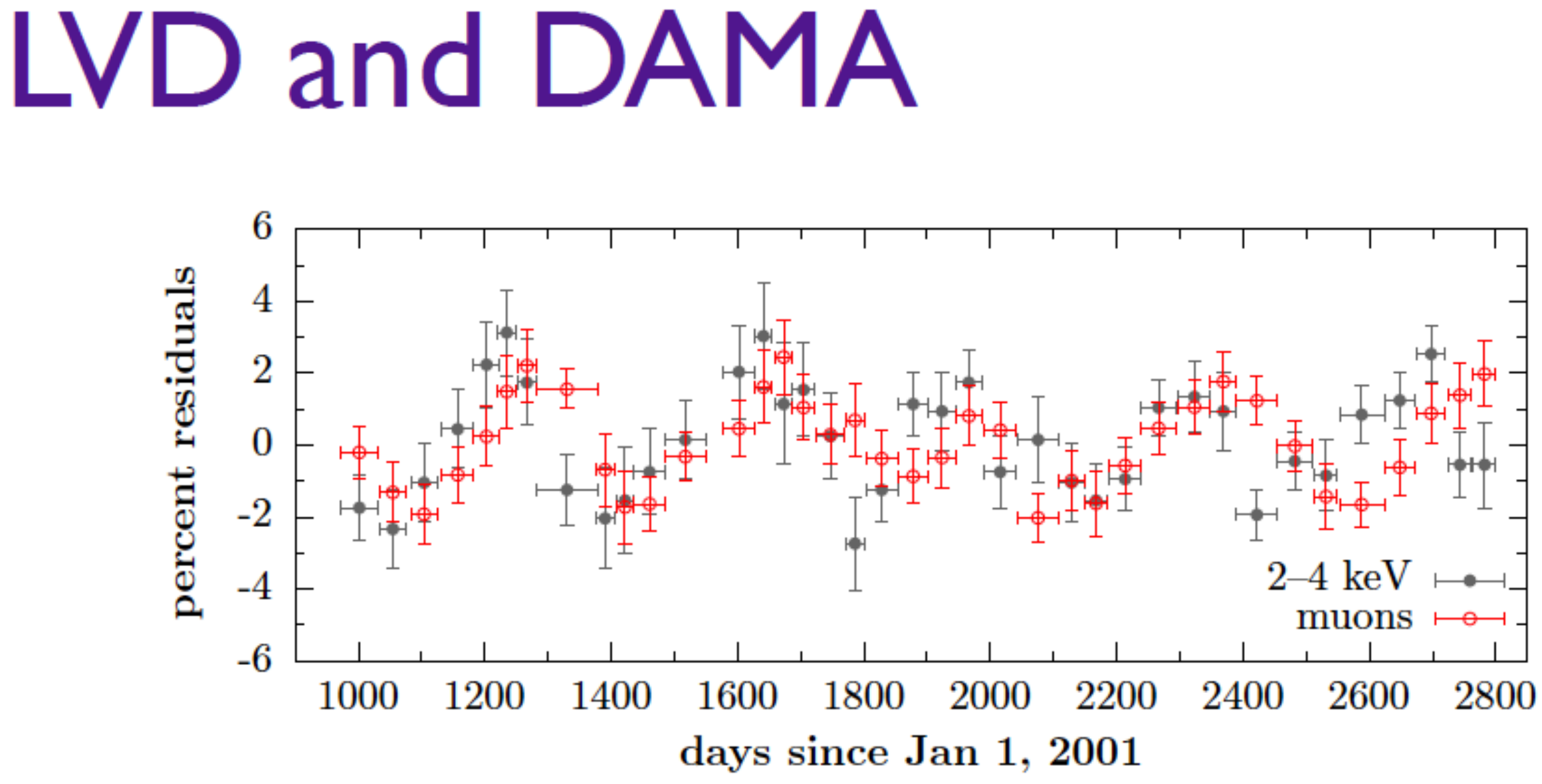

Figure 7. A comparison of the LVD muon data and the DAMA results shown at the Marina del Rey Dark Matter symposium.

## 8. New Results from CDMS

Recently at the Denver APS Meeting in March 2013 the CDMS group reported a tentative signal from their Si detectors at Low Mass (Fig. 8). These events seem to have very little ionization which is a key decrement for the CDMS WIMP search.

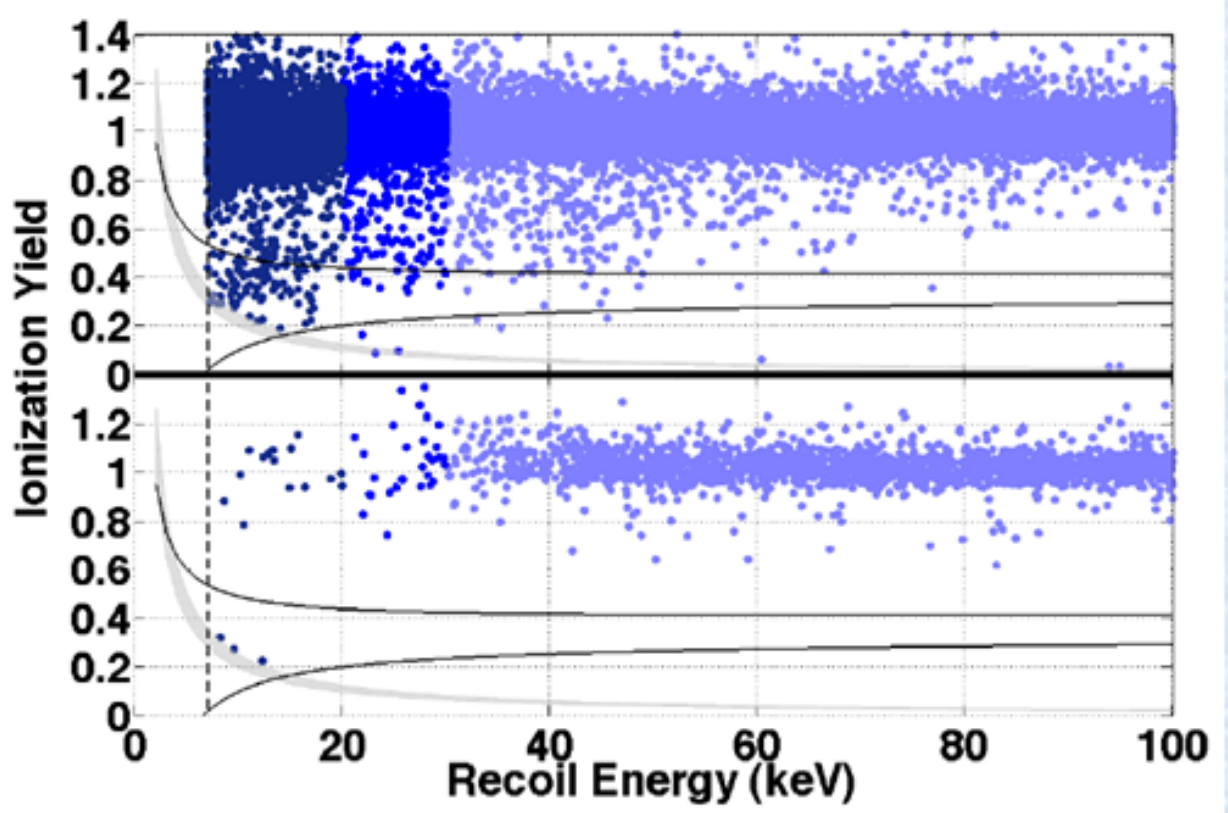

FIG. 2. Ionization yield versus recoil energy in all detectors included in this analysis for events passing all signal criteria except (*top*) and including (*bottom*) the phonon timing criterion. The curved black lines indicate the signal region (-1.8$\sigma$ and +1.2$\sigma$ from the mean nuclear recoil yield) between 7 and 100 keV recoil energies, while the gray band shows the range of charge thresholds. Electron recoils in the detector bulk have yield near unity. The data are colored to indicate recoil energy ranges (dark to light) of 7–20, 20–30, and 30–100 keV to aid the interpretation of Fig. 3.

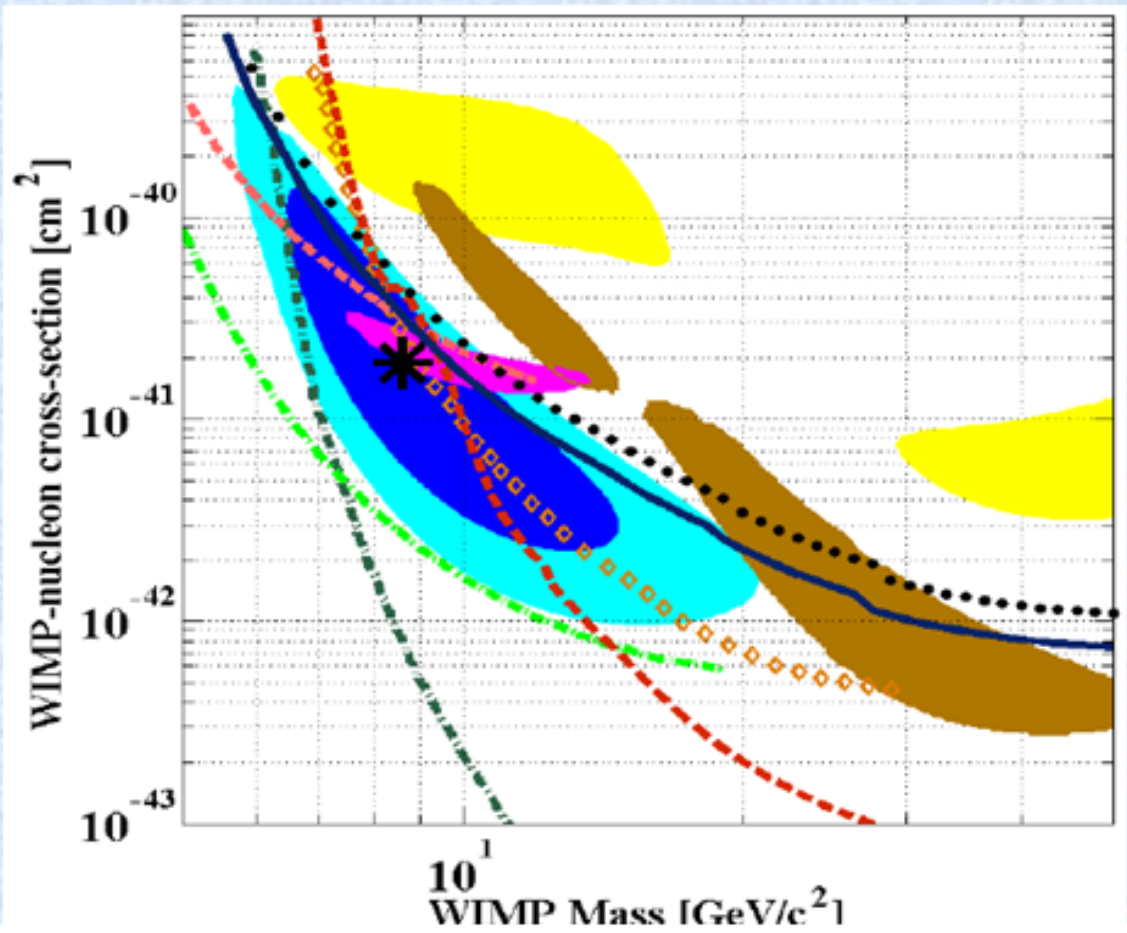

FIG. 4. Experimental upper limits (90% confidence level) for the WIMP-nucleon spin-independent cross section as a function of WIMP mass. We show the limit obtained from the exposure analyzed in this work alone (*black dots*), and combined with the CDMS II Si data set reported in [22] (*blue solid line*). Also shown are limits from the CDMS II Ge standard [11] and low-threshold [27] analysis (*dark and light dashed red*), EDELWEISS low-threshold [28] (*orange diamonds*), XENON10 S2-only [29] (*light dash-dotted green*), and XENON100 [30] (*dark dash-dotted green*). The filled regions identify possible signal regions associated with data from CoGeNT [31] (*magenta*, 90% C.L., as interpreted by Kelso *et al.* including the effect of a residual surface event contamination described in [32]), DAMA/LIBRA [16, 33] (*yellow*, 99.7% C.L.), and CRESST [18] (*brown*, 95.45% C.L.) experiments. 68% and 90% C.L. contours for a possible signal from these data are shown in blue and cyan, respectively. The asterisk shows the maximum likelihood point at (8.6 GeV/c$^2$, 1.9 $\times$ $10^{-41}$ cm$^2$).

Figure 8.

There has been a calculation by Peter Sorenson of the amount of charge that is produced by Low Mass WIMPs (Table 1) for Ge or Xe recoils (recall that the $S_2$ variable is very sensitive to even a few-electron charge. We expect the Si to produce more charge than Ge). The lack of an ionization signal on the CDMS events calls these events into question.

## $M_x$ = 8 GeV

Ge with $E_R$ = 8 KeV DM

Xe with $E_R$ = 5 KeV    ν = 709 km/sec

ν ~ 720 km/sec

Ge Atom $E_R$ = 8 KeV    $587e^-$

Xe Atom $E_R$ = 5 KeV    $21\text{-}29e^-$

2-3pe

From Peter Sorenson (private communication)

Table 1.

### 9. Use of Information Theory to Analyze XENON Data (100 Days)

Recently a group has reanalyzed the 100 day XENON 100 data (Fig 9).[14] The constraints on the low mass region are considerably better. The group used the two Pe data. As we showed before, a small $S_1$ and robust $S_2$ can reach very low mass in the Xenon detectors (Section 2). The new limit excludes the CDMS 3 event signal region.

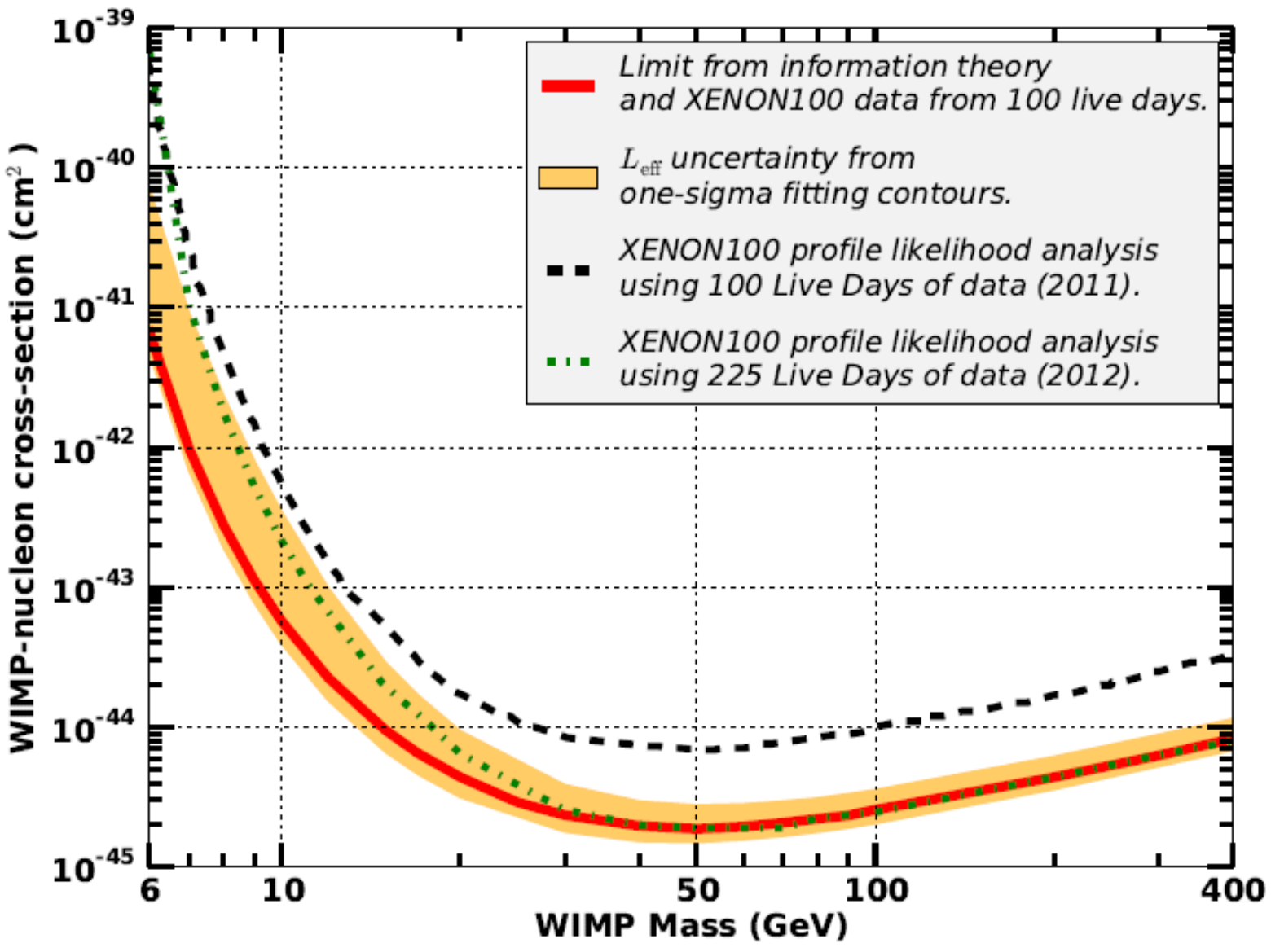

Figure 9: Limit calculated using the method presented in this letter and 100 live days of XENON100 data (solid red line) with uncertainties due to $L_{eff}$ fitting as the shaded region. The 2011 and 2012 exclusion limits obtained by the XENON100 collaboration are shown as a black dashed line and green dot-dashed line respectively.

## 9. Conclusion

We have described the claimed Low WIMP mass signals and arguments against the reality of these signals being due to WIMPs. We consider only two here:

1. The DAMA results on Annual Variation V should be stated as:

$$V_{annual} = \frac{\textit{Annual Variation signal}}{\textit{Fraction of Data that Contains WIMPs}}$$

   J. Pradler et al [12][11] have shown in Fig. 3 that the amount of WIMP signal is very small or zero. In this case $V_{annual}$ would be very large outside the region of all WIMP theories. We showed that neutron flux at the LGNS follows exactly the same time variation as DAMA.

2. The new CDMS signal sees events with very low ionizations. Calculations of WIMP interactions for an 8 GeV WIMP by Sorenson indicate an expected large ionization. There is a conflict making it unlikely that the CDMS events are due to WIMPs. A new analysis of XENON100 data further excludes the CDMS II signal region.

I wish to thank Peter Sorenson for discussions and the XENON100 group. Some of this work was done at the Aspen Institute for Physics.